\documentclass[pra,a4paper,showpacs,superscriptaddress]{revtex4}
\usepackage{amsmath}
\usepackage{amsfonts}
\usepackage{graphicx}
\usepackage{longtable}
\newcommand{\be}{\begin{equation}}
\newcommand{\ee}{\end{equation}}

\newcommand{\ba}[1]{\left(\begin{array}{#1}}
\newcommand{\ea}{\end{array}\right)}
\begin{document}

\title{Bipartite separability of one parameter families of states using conditional quantum relative Tsallis entropy}
\author{ Anantha S Nayak}
\affiliation{Department of Physics, Kuvempu University, 
Shankaraghatta, Shimoga-577 451, India} 
\author{Sudha }
\affiliation{Department of Physics, Kuvempu University, 
Shankaraghatta, Shimoga-577 451, India}
\affiliation{Inspire Institute Inc., Alexandria, Virginia, 22303, USA.}
\author{A. K. Rajagopa1}\affiliation{Inspire Institute Inc., Alexandria, Virginia, 22303, USA.}
\affiliation{Harish-Chandra Research Institute, Chhatnag Road, Jhunsi, Allahabad 211 019, India}
\affiliation{Institute of Mathematical Sciences, C.I.T. Campus, Taramani, Chennai, 600113, India} 
\author{A. R. Usha Devi}
\affiliation{ Department of Physics, Bangalore University, Bangalore 560 056, India}
\affiliation{Inspire Institute Inc., Alexandria, Virginia, 22303, USA.}
\date{\today}
\begin{abstract} 
In any bipartition of a quantum state, it is proved that the negative values of the conditional version of sandwiched Tsallis relative entropy necessarily implies quantum entanglement. For any $N$, the separability ranges in the $1:N-1$ partition of symmetric one parameter families of noisy $N$-qubit $W$-~, GHZ-, $W\bar{W}$ states are determined using the conditional quantum relative Tsallis entropy approach.  The $1:N-1$ separability range matches exactly with the range obtained through positive partial transpose criterion, for all $N$. The advantages of using non-commuting version of $q$-conditional relative Tsallis entropy is brought out through this and other one-parameter families of states.       
\end{abstract}
\maketitle 
\section{Introduction}
\label{intro}
It is well known that entropic characterization of separability captures the local versus global disorder of mixed composite states and serves as a convenient tool in identifying bipartite separability range in several one-parameter families of states $\rho(x)$~\cite{ent,hki,aber,tsa,sabe,canosa,jb,prabhu,arss,asanu}.  The quantum versions of more generalized entropies such as R\'{e}nyi and Tsallis entropies are found to yield better separability range than those obtained through von-Neumann entropy~\cite{aber,tsa,sabe,canosa,jb,prabhu}. In fact, positivity of Tsallis conditional entropy is found to capture global vs local disorder in mixed states much better than that through von-Neumann conditional entropy leading to  stricter range of separability~\cite{aber,tsa,sabe,canosa,jb,prabhu,arss}. 

Making use of the quantum generalization of  R\'{e}nyi relative entropy to the situation when the pair of density matrices are noncommuting~\cite{mw,mds,renyimax} , 
an analogous generalization to Tsallis relative entropy and its conditional version are obtained in Ref.~\cite{asanu}. This generalization was done in anticipation of the fact that the conditional version of generalized Tsallis relative entropy is more effective in identifying entangled states than its traditional commuting version.     

The so-called `sandwiched' Tsallis relative entropy~\cite{asanu} is given by
\be
\label{gstre}
\tilde{D}_q^{T}(\rho\vert\vert \sigma)=\frac{\mbox{Tr}\left\{\left(\sigma^{\frac{1-q}{2q}}\rho \, \sigma^{\frac{1-q}{2q}}\right)^q \right\}-1}{q-1}
\ee
quite on the same lines of the definition of generalized version of quantum relative R\'{e}nyi entropy~\cite{mw,mds,renyimax}. 
It reduces to the traditional relative Tsallis entropy 
\be
D_q^{T}(\rho\vert\vert\sigma)=\frac{\mbox{Tr}\left( \rho^q \sigma^{1-q} \right)-1}{q-1},
\ee
when $\sigma$ and $\rho$ commute with each other. It is useful here to note the Lieb-Thirring inequality $\tilde{D}_q^{T}(\rho\vert\vert \sigma)\leq D_q^{T}(\rho\vert\vert\sigma)$.     

The conditional version of $\tilde{D}_q^{T}(\rho\vert\vert \sigma)$ is defined as~\cite{asanu} 
\begin{eqnarray}
\label{cstre1}
\tilde{D}^{T}_q(\rho_{AB}||\rho_B)&=&\frac{\tilde{Q}_q(\rho_{AB}||\rho_B)-1}{1-q} 
\end{eqnarray}
where 
\be
\label{qab}
\tilde{Q}_q(\rho_{AB}||\rho_B)=\mbox{Tr}\left\{\left[(I_A\otimes \rho_B)^{\frac{1-q}{2q}}\rho_{AB}(I_A\otimes \rho_B)^{\frac{1-q}{2q}}\right]^q\right\}.
\ee
In fact, $\tilde{D}^{T}_q(\rho_{AB}||\rho_B)$ reduces to the  Abe-Rajagopal(AR) $q$-conditional Tsallis entropy~\cite{aber}  
\be 
\label{ar}
 S_q^{T}(A\vert B)=\frac{1}{1-q}\left(\frac{\mbox{Tr}\rho_{AB}^q}{\mbox{Tr}\rho_B^q}-1 \right)
\ee
when the subsystem density matrix $\rho_B$ is a maximally mixed state, thus commuting with 
$\rho_{AB}$. The AR $q$-conditional entropy $S_q^{T}(A\vert B)$ is known to reduce to the von-Neumann conditional entropy in the limit $q\rightarrow 1$~\cite{aber}.  Making use of the fact that $S_q^{T}(A\vert B)<0$ for entangled states, $S_q^{T}(A\vert B)$ is found to be more efficient than von-Neumann conditional entropy in detecting entangled states in the limit $q\rightarrow \infty$~\cite{aber}.      
The AR criterion has been employed to examine separability of several classes of composite 
states~\cite{aber,tsa,sabe,canosa,jb,prabhu,arss}. Though the AR criterion provides a better 
separability criterion than the one using von-Neumann conditional 
entropy~\cite{aber,sabe,prabhu}, in some cases~\cite{prabhu} the 
separability range obtained through AR criterion is found to be larger than 
that obtained through Peres' PPT criterion~\cite{peres}, thus making AR-criterion 
weaker in comparison with PPT criterion.  A closer examination revealed that\cite{asanu} 
such cases correspond to the situations where the subsystem density matrix  $\rho_B$ is 
not maximally mixed thereby not commuting with its original density matrix $\rho_{AB}$. An 
entropic criterion which takes into account the aspect of non-commutativity of subsystem 
density matrices with their original density matrices was thus found necessary.  
This aspect has been addressed in Ref.~\cite{asanu} using the 
{\emph {non-commuting version}} 
of Tsallis relative entropy and its conditional version. 

The criterion which makes use of the Conditional version of Sandwiched 
Tsallis Relative Entropy (CSTRE), the so-called CSTRE criterion, detects entanglement in a state 
using negative values of the quantity $\tilde{D}^{T}_q(\rho_{AB}||\rho_B)=\frac{\tilde{Q}_q(\rho_{AB}||\rho_B)-1}{1-q}$ in the limit $q\rightarrow \infty$, quite similar to AR-criterion. As $\tilde{D}^{T}_q(\rho_{AB}||\rho_B)$ reduces to AR q-conditional entropy in situations where $\rho_B$ is maximally mixed and hence commuting with $\rho_{AB}$, the results of AR- and CSTRE-criterion match with each other whenever $\rho_{B}$ corresponds to a maximally mixed state.  
The CSTRE criterion has been employed to identify the separability range in the $3$-, $4$- qubit one parameter families of noisy W-, GHZ- states in Ref.~\cite{asanu}.  Stricter separability range than that obtained through AR $q$-conditional entropy~\cite{aber} is seen to be achievable, in the case of one-parameter family of noisy W states~\cite{asanu}, using the CSTRE criterion. Also, in the $1:2$ and $1:3$ partitions of $3$-, $4$- qubit one-parameter families of W-, GHZ- states, the separability range is seen to match exactly~\cite{asanu} with that through Peres' PPT criterion~\cite{peres}. Our aim here is to identify the $1:N-1$ separability ranges in the symmetric one-parameter families of $N$-qubit noisy mixed states using the CSTRE criterion and examine whether it matches with the results of PPT criterion. We also examine several different one-parameter families of states and illustrate the utility of CSTRE criterion in identifying bipartite entangled states.  

The article is organized as under: Section \ref{intro} gives a brief review of the entropic separability criteria, defines the conditional version of sandwiched relative Tsallis entropy as the non-commuting generalization of AR-q conditional entropy criterion and details the motivation for the present work. In Section~\ref{sec2} we prove an important theorem authenticating the use of CSTRE in detecting quantum entanglement in any bipartite quantum state.  In Section~\ref{sec3}, divided into three subsections, we obtain the {\emph{separability range in the $1:N-1$ partition of symmetric $N$-qubit noisy states using the CSTRE criterion}} and show that they match exactly with that through PPT criterion (Secs. 3.1, 3.2).  Sec. 3.3 discusses the use of conditional version of Renyi relative entropy in identifying entangled states and contains an account of other types of states that can be dealt with using CSTRE criterion. Section~\ref{sec4} gives a summary of the work as well as future problems of interest in this framework.  

\section{Sufficient condition for quantum entanglement in terms of conditional version of sandwiched Tsallis relative entropy} 
\label{sec2} 
Before proceeding to make use of the non-commuting version of Tsallis relative entropy $\tilde{D}^{T}_q(\rho_{AB}||\rho_B)$ for $N$-qubit states, we wish to establish that `{\emph{negative values of $\tilde{D}^{T}_q(\rho_{AB}||\rho_B)$ in any bipartition of a composite state $\rho_{AB}$ indicates entanglement in that bipartition}'. Our use of the conditional version of `sandwiched'(non-commuting) Tsallis relative entropy, in identifying entanglement in bipartite quantum states, is thus best exemplified through the following theorem.

\noindent{{\large{\bf{Theorem:}}}  {\emph{Negative values of the conditional version of the sandwiched Tsallis relative entropy (CSTRE) $\tilde{D}^{T}_q(\rho_{AB}||\rho_B)$ with $q>1$ {\bf{necessarily imply entanglement}} in the state $\rho_{AB}$}}. 

\noindent{\bf{Proof:}} 
We know that for any two positive semi-definite operators $\rho$ and $\sigma$, the trace functional~\cite{ndt}    $\tilde{Q}_q(\rho||\sigma)=\mbox{Tr}\left\{\left[\sigma^{\frac{1-q}{2q}}\rho \sigma^{\frac{1-q}{2q}}\right]^q\right\}$ satisfies the inequality~\cite{ndt} 
\be
\label{oineq} 
\tilde{Q}_q(\rho||\sigma)\leq \tilde{Q}_q(\rho||\rho)\ \  \mbox{for} \ \ q>1 \ \ \mbox{whenever} \ \ \rho \leq \sigma. 
\ee 
Notice that when $\rho$ is a density matrix, $\tilde{Q}_q(\rho||\rho)=\mbox{Tr}\,\rho=1$ implying that 
\be
\label{oineq2} 
\tilde{Q}_q(\rho||\sigma) \leq 1 \ \ \mbox{when} \ \ \rho \leq \sigma \ \  \mbox{and} \ \ q>1.
\ee 
With $\rho=\rho_{AB}$, $\sigma=I_A\otimes \rho_B$ and $\tilde{Q}_q(\rho_{AB}||I_A \otimes \rho_B)\equiv \tilde{Q}_q(\rho_{AB}||\rho_B)$,  Eq. (\ref{oineq2}) gives 
\begin{eqnarray}
\label{oineq3}
& & \tilde{Q}_q(\rho_{AB}||\rho_B)=\mbox{Tr}\left\{\left[(I_A\otimes \rho_B)^{\frac{1-q}{2q}}\rho_{AB}(I_A\otimes \rho_B)^{\frac{1-q}{2q}}\right]^q\right\} \leq 1   \\
& & \mbox{when} \ \ \rho_{AB} \leq I_A\otimes \rho_B \ \  \mbox{and} \ \ q>1. \nonumber
\end{eqnarray}   
We now recall that, for all separable states $\rho_{AB}$,
\be 
\label{rc}
\rho_{AB}-\left(I_A \otimes \rho_{B}\right)\leq 0   
\ee   
according to reduction criterion~\cite{rdc}.
Thus, as  $\rho_{AB}\leq I_A \otimes \rho_{B}$ for all separable states $\rho_{AB}$, we have 
\be 
\label{oineq3} 
\tilde{Q}_q(\rho_{AB}||\rho_B)\leq 1 \ \ \mbox{whenever} \ \ q>1. 
\ee
It can now be readily seen that $\tilde{D}^{T}_q(\rho_{AB}||\rho_B)=\frac{\tilde{Q}_q(\rho_{AB}||\rho_B)-1}{1-q}$ with $q>1$ is non-negative for all separable states. In other words, negative values of the conditional version of sandwiched Tsallis relative entropy (CSTRE) $\tilde{D}^{T}_q(\rho_{AB}||\rho_B)$ ($q>1$) indicate entanglement in the state $\rho_{AB}$ thus proving the theorem.   

Through Theorem~1, we have established that `{\emph{negativity}}' of CSTRE ($\tilde{D}^{T}_q(\rho_{AB}||\rho_B)$ ($q>1$)) is a `{\emph{sufficient criterion}}' for the bipartite state $\rho_{AB}$ {\emph{to be entangled}}. We make use of this fact and in one-parameter family of symmetric states we identify the value of the parameter $x$  at which $\tilde{D}^{T}_q(\rho_{AB}||\rho_B)$ ($q>1$) changes from positive to negative or vice versa in the limit $q\rightarrow \infty$. In other words, we identify the `{\emph{zero(s)}}' of $\tilde{D}^{T}_q(\rho_{AB}||\rho_B)$ when $q\rightarrow \infty$ and separability range(s) of the state $\rho_{AB}$ correspond to the range(s) of the parameter $x$ in which $\tilde{D}^{T}_q(\rho_{AB}||\rho_B)\geq 0$ in the limit $q\rightarrow \infty$.

\section{One parameter families of symmetric $N$-qubit mixed states} 
\label{sec3}
The symmetric one parameter family of noisy $N$-qubit mixed states is given by  
\be
\label{noisy} 
\rho_N(x)=\left(\frac{1-x}{N+1}\right)P_N  + x\vert \Phi_N \rangle \langle \Phi_N \vert
\ee
with $P_N=\sum_{M=-\frac{N}{2}}^{\frac{N}{2}}\,\left\vert \frac{N}{2},\,M \right\rangle \left\langle \frac{N}{2},\,M\right\vert$ is the projector 
onto the $N+1$ dimensional maximal multiplicity subspace of the collective angular momentum of $N$-qubits, 
$\vert \frac{N}{2},\,M \rangle$ being the basis states of this subspace. $\vert \Phi_N\rangle$ is any pure state belonging to this symmetric subspace. Notice that $x$ is a parameter lying in the range $[0,\,1]$ and when $x=0$ the state $\rho_N(x)$ is maximally mixed (in the symmetric subspace) whereas it is a pure state $\vert \Phi_N \rangle$ when $x=1$. The separability range of one parameter family of mixed states refers to the range of values $x$ in which the state $\rho_N(x)$ is separable, in a chosen bipartition of the state. The separability ranges differ for each bipartition and in this work we analyze the $1:N-1$ bipartition of the state $\rho_N(x)$.  
\subsection{One-parameter family of noisy W states}
We have the symmetric one-parameter family of noisy  W states  
\begin{equation}
\label{nw}
\rho_N^{(W)}(x)=\left(\frac{1-x}{N+1}\right)P_N+ x\vert W_N \rangle\langle W_N \vert
\end{equation} 
where $\vert W_N \rangle\equiv \left\vert \frac{N}{2},\,\frac{N}{2}-1 \right\rangle$ is one among the basis states of the $N+1$ dimensional symmetric subspace of collective angular momentum. 
We recall here that using the Abe-Rajagopal $q$-conditional entropy~\cite{aber}, the separability range of the $3$-qubit state $\rho_3^{(W)}(x)$, in its $1:2$ partition, is found to be $[0,\,0.2]$ while the PPT criterion gives the stricter separability range $[0,\,0.1547]$~\cite{prabhu,asanu}. In the $1:3$ partition of the 4-qubit state $\rho_4^{(W)}(x)$ also, the AR criterion leads to the weaker separability range $[0,\,0.1666]$ compared to the range $[0,\,0.1123]$ obtained through PPT criterion. An observation of the fact that the single qubit density matrix of $\rho_N^{(W)}(x)$ is not maximally mixed led Rajagopal et.al.,~\cite{asanu} to make use of the non-commuting version of the Tsallis relative entropy~\cite{mw,mds} to obtain a better separability range for the case under examination. They proposed the conditional version of the sandwiched Tsallis relative entropy~\cite{asanu} (CSTRE) $\tilde{D}^{T}_q(\rho_{AB}||\rho_B)$ and examined the range in which it is negative in the limit $q\rightarrow \infty$. Quite in accordance with the expectations, the CSTRE criterion resulted in a better separability range~\cite{asanu} than that through AR criterion and it even matched with the $1:2$, $1:3$  separability ranges of $\rho_3^{(W)}(x)$, $\rho_4^{(W)}(x)$ obtained through PPT criterion. 

Continuing further with the use of AR criterion, the $1: N-1$ separability range of the one-parameter family of noisy W states has been obtained in Ref. \cite{prabhu} and it is found to be  
\be
\label{abW}
0\leq x < \frac{1}{N+2}
\ee
for any $N\geq 3$~\cite{prabhu}. This has been a generalization of their result for $\rho_3^{(W)}(x)$, $\rho_4^{(W)}(x)$, in their respective $1:N-1$ partitions, to $\rho_N^{(W)}(x)$. As we have seen that~\cite{asanu} for $3$- and $4$-qubit noisy W states the CSTRE criterion yields a stricter separability range than that obtained through AR criterion, our immediate interest is to generalize the CSTRE separability range to $N$-qubit states $\rho_N^{(W)}(x)$, in their $1:N-1$ partition, for any $N\geq 3$. We carry that out in the following. 

In order to find the $1:N-1$ separability range of the state $\rho_N^{(W)}(x)$, we need to evaluate the eigenvalues $\lambda_i$ of the `sandwiched' matrix 
$(I_A\otimes \rho_B)^{\frac{1-q}{2q}}\rho_N^{(W)}(x)(I_A \otimes \rho_B)^{\frac{1-q}{2q}}$, ($\rho_B=\mbox{Tr}_A\,\rho_{AB}$ being the subsystem density matrix of $\rho_{AB}$ corresponding to $N-1$ qubits)  so that (See Eq. (\ref{qab}))  $\tilde{Q}_q(\rho_N^{(W)}(x)||\rho_B)=\sum_{i=1}^{N+1} \,\lambda_i^q$ and 
\begin{eqnarray} 
\label{cstre2}
\tilde{D}^{T}_q(\rho_N^{(W)}(x)||\rho_B)&=& 
                                  \frac{\sum_{i=1}^{N+1} \,\lambda_i^q-1}{1-q}.  
\end{eqnarray}
Here, as our interest is to find out the $1:N-1$ separability range we have taken the subsystems $A$, $B$ to correspond respectively to a {\emph{single qubit}} and the remaining $N-1$ qubits (the state $\rho_N^{(W)}(x)$ is symmetric and it does not matter which qubit we take as subsystem $A$). {\emph{According to CSTRE criterion, the $1:N-1$ separability range of $\rho_N^{(W)}(x)$ is the range in which the parameter $x$  gives non-negative values for $\tilde{D}^{T}_q(\rho_N^{(W)}(x)||\rho_B)$, in the limit $q\rightarrow \infty$.}} 

The non-zero eigenvalues $\lambda_i$, $i=1,\,2,\ldots,N+1$ being crucial in the evaluation of $\tilde{D}^{T}_q(\rho_N^{(W)}(x)||\rho_B)$, we examine the form of these eigenvalues when $N=3, 4, 5, 6$ to analyze whether a generalization to the case of any $N$ is possible. We explicitly evaluate the eigenvalues $\lambda_i$  of the sandwiched matrix $(I_A\otimes \rho_B)^{\frac{1-q}{2q}}\rho_N^{(W)}(x)(I_A \otimes \rho_B)^{\frac{1-q}{2q}}$   when $N=3,\, 4, \,5,\, 6$ and  the following table (Table~1) provides the non-zero eigenvalues. 
\begin{table}[h] 
\begin{center} 
\caption{The non-zero eigenvalues $\lambda_i$ of the sandwiched matrix $(I_A\otimes \rho_B)^{\frac{1-q}{2q}}\rho_N^{(W)}(x)(I_A \otimes \rho_B)^{\frac{1-q}{2q}}$ for $N=3\ to \ 6$}
\scriptsize{
\begin{tabular}{|c|c|c|c|c|}
\hline
& & & & \\
Number & $\lambda_1$  &  &   &    \\ 
 of & $(N-2)$ fold &  $\lambda_2$  & $\lambda_3$ & $\lambda_4$   \\
qubits ($N$) & degenerate & &  &  \\
\hline\hline 
$N=3$  & $\left(\frac{1-x}{4}\right)\left(\frac{1-x}{3}\right)^{\frac{1-q}{q}}$  & $\left(\frac{1-x}{4}\right)\left(\frac{1}{3}\right)^{\frac{1-q}{q}}$ & $\left(\frac{1-x}{4}\right)\left(\frac{1}{3}\right)^{\frac{1}{q}}\left[ 
(1-x)^{\frac{1-q}{q}}+2(1+x)^{\frac{1-q}{q}} \right] $      &  $\left(\frac{1+3x}{4}\right)\left(\frac{1}{3}\right)^{\frac{1}{q}}\left[ 
1+2(1+x)^{\frac{1-q}{q}} \right]$ \\ 
 & & & &   \\ \hline 
 $N=4$  & $\left(\frac{1-x}{5}\right)\left(\frac{1-x}{4}\right)^{\frac{1-q}{q}}$  & $\left(\frac{1-x}{5}\right)\left(\frac{1}{4}\right)^{\frac{1-q}{q}}$ & $\left(\frac{1-x}{5}\right)\left(\frac{1}{4}\right)^{\frac{1}{q}}\left[ 
2(1-x)^{\frac{1-q}{q}}+2(1+2x)^{\frac{1-q}{q}} \right] $      &  $\left(\frac{1+4x}{5}\right)\left(\frac{1}{4}\right)^{\frac{1}{q}}\left[ 
1+3(1+2x)^{\frac{1-q}{q}} \right]$ \\ 
 & & & &   \\ \hline 
$N=5$  & $\left(\frac{1-x}{6}\right)\left(\frac{1-x}{5}\right)^{\frac{1-q}{q}}$  & $\left(\frac{1-x}{6}\right)\left(\frac{1}{5}\right)^{\frac{1-q}{q}}$ & $\left(\frac{1-x}{6}\right)\left(\frac{1}{5}\right)^{\frac{1}{q}}\left[ 
3(1-x)^{\frac{1-q}{q}}+2(1+3x)^{\frac{1-q}{q}} \right] $      &  $\left(\frac{1+5x}{6}\right)\left(\frac{1}{5}\right)^{\frac{1}{q}}\left[ 
1+4(1+3x)^{\frac{1-q}{q}} \right]$ \\ 
 & & & &   \\ \hline 
$N=6$  & $\left(\frac{1-x}{7}\right)\left(\frac{1-x}{6}\right)^{\frac{1-q}{q}}$  & $\left(\frac{1-x}{7}\right)\left(\frac{1}{6}\right)^{\frac{1-q}{q}}$ & $\left(\frac{1-x}{7}\right)\left(\frac{1}{6}\right)^{\frac{1}{q}}\left[ 
4(1-x)^{\frac{1-q}{q}}+2(1+4x)^{\frac{1-q}{q}} \right] $      &  $\left(\frac{1+6x}{7}\right)\left(\frac{1}{6}\right)^{\frac{1}{q}}\left[ 
1+5(1+4x)^{\frac{1-q}{q}} \right]$ \\ 
 & & & &   \\ \hline 
\end{tabular}}
\end{center}
\end{table} 

An explicit evaluation of the four eigenvalues for arbitrary $N$ is tedious and so we notice the trends of each eigenvalue presented in the columns of Table~1 with increasing $N=3,\,4,\,5,\,6$. This clearly suggests the nature of eigenvalues for any $N$ and they are given below.

\begin{eqnarray}
\label{geneigw}
\lambda_1&=&\left(\frac{1-x}{N+1}\right)\left(\frac{1-x}{N}\right)^{\frac{1-q}{q}},\ \  (N-2)\  \mbox{fold degenerate}; \nonumber \\
\lambda_2&=&\left(\frac{1-x}{N+1}\right)\left(\frac{1}{N}\right)^{\frac{1-q}{q}}, \nonumber \\
\lambda_3&=&\left(\frac{1-x}{N+1}\right) \left(\frac{1}{N}\right)^{\frac{1}{q}} \left[(N-2)(1-x)^\frac{1-q}{q}+2(1+(N-2)x)^{\frac{1-q}{q}}\right], \\
\lambda_4&=&\left(\frac{1+Nx}{N+1}\right) \left(\frac{1}{N}\right)^{\frac{1}{q}}\left[1 + (N-1)\left(1 + (N-2)x\right)^\frac{1-q}{q}\right].\nonumber 
\end{eqnarray} 
On making use of the values of $\lambda_i$, $i=1,\,2,\,3,\,4$, we identify the zero(s) of CSTRE $\tilde{D}^{T}_q(\rho_N^{(W)}(x)||\rho_B)$ (See Eq. (\ref{cstre2})) for any $N$ in the limit $q\rightarrow \infty$. It can be seen that $\tilde{D}^{T}_q(\rho_N^{(W)}(x)||\rho_B)$ is a monotonically decreasing function for all values of $N$ and $q>1$. In the limit $q\rightarrow \infty$, the only zero of $\tilde{D}^{T}_q(\rho_N^{(W)}(x)||\rho_B)$ occurs at the value $x=\frac{-N+\sqrt{2N(N-1)}}{N(N-2)}$. Thus, the separability range obtained through CSTRE, in the $1:N-1$ partition, of the state $\rho_N^{(W)}(x)$ ($N\geq 3$) is found to be  
\be
\label{sepw}
0 \leq x \leq \frac{-N+\sqrt{2N(N-1)}}{N(N-2)}. 
\ee
Note that Eq. (\ref{sepw}) is different from the AR result in Eq. (\ref{abW}). {\emph{We assert here that it is the non-commutativity of the single qubit marginal 
given by $\rho_1=\rho_2=\cdots \rho_N=\mbox{diag}\left(\frac{N+(N-2)x}{2N}, \, \frac{N-(N-2)x}{2N} \right)$ with $\rho_N^{(W)}(x)$ that results in a stricter separability range through CSTRE criterion compared to its commuting version, the AR criterion}}.  
Also one can immediately recover the range 
$(0,\,0.1547)$, $(0,\,0.1123)$ respectively for the states $\rho_3^{(W)}(x)$, $\rho_4^{(W)}(x)$ using the relation Eq. (\ref{sepw}) and this is in accordance with the range obtained using the CSTRE criterion directly for the $3$-, $4$- qubit states $\rho_3^{(W)}(x)$, $\rho_4^{(W)}(x)$ in Ref. \cite{asanu}. We also obtain the  separability ranges $(0,\,0.0883)$, $(0,\,0.07275)$ in the    
$1:4$, $1:5$ partitions respectively for the $5$- and $6$- qubit states of the family of noisy W states. We have verified that these separability ranges (for $N=3,\,4,\,5,\,6$) match with those obtained through PPT criterion. We can thus conjecture that the CSTRE separability range in Eq. (\ref{sepw}) for the $1:N-1$ partition of the states $\rho_N^{(W)}(x)$  is also the PPT separability range.  

\begin{figure}[ht]
\begin{center}
\includegraphics* [width=2.4in,keepaspectratio]{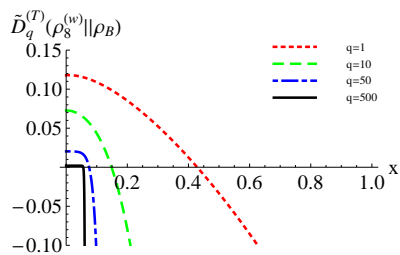} 
\caption{(Color Online)The conditional form of sandwiched Tsallis relative entropy $\tilde{D}^{T}_q(\rho^{(W)}_8(x)\vert\vert \rho_{B})$ for one-parameter family of $8$-qubit W-states  as a function of $x$ for different values of $q$. It can be seen that $\tilde{D}^{T}_q(\rho^{(W)}_8(x)\vert\vert \rho_{B})$  is negative for $x>0.4246$ when $q=1$(separability range through von-Neumann conditional entropy), whereas it is negative for  $x>0.0538$ in the limit $q\rightarrow \infty$ (separability range through CSTRE criterion) . Observe that the range of values of $x$ for which $\tilde{D}^{T}_q(\rho^{(W)}_8(x)\vert\vert \rho_{B})$ is non-negative becomes smaller as $q>1$ becomes larger and is the smallest when $q\rightarrow \infty$. All the quantities are dimensionless.} 
\end{center}
\end{figure} 

\begin{figure}[ht]
\begin{center}
\includegraphics* [width=2.4in,keepaspectratio]{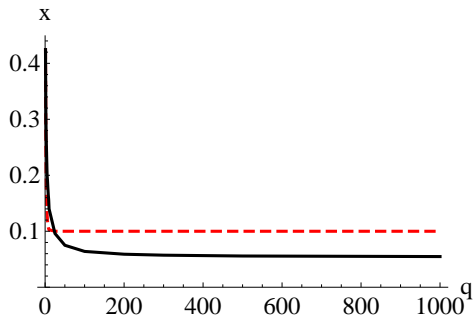} 
\caption{(Color Online) Implicit plot of $\tilde{D}^{T}_q(\rho^{(W)}_8\vert\vert \rho_{B})=0$ as a function of $q$ (solid line) indicating that  
$x\rightarrow 0.0538$ as $q\rightarrow\infty$. In contrast, the implicit plot of Abe-Rajagopal $q$-conditional entropy $S_q^{T}(A\vert B)=0$ (dashed line) leads to $x\rightarrow 0.1$ as $q\rightarrow\infty$. The quantities plotted are dimensionless.}  
\end{center}
\end{figure}

In Fig. 1 we have illustrated the nature of variation of $\tilde{D}^{T}_q(\rho^{(W)}_8(x)\vert\vert \rho_{B})$ with $q$ as well as $x$ and the identification of the $1:7$ separability range for the one-parameter family of $8$-qubit noisy W state $\rho_8^{(W)}(x)$. It is worth observing that the range of $x$ in which $\tilde{D}^{T}_q(\rho^{(W)}_8(x)\vert\vert \rho_{B})$ is non-negative becomes smaller as $q$ increases. In fact, at $q=1$, we get the separability range through von-Neumann conditional entropy, which is much larger than the separability range obtained through CSTRE criterion in the limit $q\rightarrow \infty$.    
Fig. 2 compares the separability range obtained through AR criterion with the one through CSTRE criterion.

From Eq. (\ref{sepw}) and the discussion following it, it can be readily seen that the $1:N-1$ separability range of $\rho_N^{(W)}(x)$ reduces considerably with the increase in $N$. Thus, for large $N$ (macroscopic limit), one can expect that  a single qubit and its remaining $N-1$ qubits are entangled for the whole range $0\leq x \leq 1$ in the state $\rho_N^{(W)}(x)$.   
\subsection{One-parameter family of noisy GHZ states} 
We now examine the one-parameter family of noisy  GHZ states 
\begin{equation}
\label{nghz}
\rho_N^{(GHZ)}(x)=\left(\frac{1-x}{N+1}\right)P_N+ x\vert GHZ \rangle_N \langle GHZ \vert
\end{equation}
in order to find their $1:N-1$ separability range, using CSTRE criterion. In fact, in Ref~\cite{asanu} it has been shown that for $3$- and $4$-qubit states $\rho_N^{(GHZ)}(x)$,  their respective $1:N-1$ separability ranges obtained using CSTRE criterion matched exactly with that through AR  and PPT criteria. We now wish to generalize this result to $N$-qubit states $\rho_N^{(GHZ)}(x)$ and our method is similar to the one adopted for one parameter family of noisy W states in the previous section. 

The eigenvalues $\mu_i$ of the sandwiched matrix $(I_A\otimes \rho_B)^{\frac{1-q}{2q}}\rho_N^{(GHZ)}(x)(I_A \otimes \rho_B)^{\frac{1-q}{2q}}$ for $N=3$ to $6$ are given in Table 2. 
 \begin{table}[h] 
\begin{center} 
\caption{The eigenvalues $\mu_i$ of the sandwiched matrix $(I_A\otimes \rho_B)^{\frac{1-q}{2q}}\rho_N^{(GHZ)}(x)(I_A \otimes \rho_B)^{\frac{1-q}{2q}}$ for $N=3,\, 4,\,5,\,6$}
\scriptsize{
\begin{tabular}{|c|c|c|c|c|}
\hline
& & & & \\
Number & $\mu_1$  &  &   & $\mu_4$   \\ 
 of & $(N-3)$-fold &  $\mu_2$  & $\mu_3$ & $2$-fold   \\
qubits ($N$) & degenerate & &  & degenerate  \\
\hline\hline 
$N=3$  & --  & $\left(\frac{1-x}{4}\right)\left(\frac{2+x}{6}\right)^{\frac{1-q}{q}}$ & $\left(\frac{1+3x}{4}\right)\left(\frac{2+x}{6}\right)^{\frac{1-q}{q}}$  &  $\left(\frac{1-x}{4}\right)\left(\frac{1}{3}\right)^{\frac{1}{q}}\left[ 
2(1-x)^{\frac{1-q}{q}}+(1+x/2)^{\frac{1-q}{q}} \right]$ \\ 
 & & & &   \\ \hline 
$N=4$  & $\left(\frac{1-x}{5}\right)\left(\frac{1-x}{4}\right)^{\frac{1-q}{q}}$  & $\left(\frac{1-x}{5}\right)\left(\frac{1+x}{4}\right)^{\frac{1-q}{q}}$ & $\left(\frac{1+4x}{5}\right)\left(\frac{1+x}{4}\right)^{\frac{1-q}{q}}$  &  $\left(\frac{1-x}{5}\right)\left(\frac{1}{4}\right)^{\frac{1}{q}}\left[ 
3(1-x)^{\frac{1-q}{q}}+(1+x)^{\frac{1-q}{q}} \right]$ \\  
 & & & &   \\ \hline 
$N=5$  & $\left(\frac{1-x}{6}\right)\left(\frac{1-x}{5}\right)^{\frac{1-q}{q}}$  & $\left(\frac{1-x}{6}\right)\left(\frac{2+3x}{10}\right)^{\frac{1-q}{q}}$ & $\left(\frac{1+5x}{6}\right)\left(\frac{2+3x}{10}\right)^{\frac{1-q}{q}}$  &  $\left(\frac{1-x}{6}\right)\left(\frac{1}{5}\right)^{\frac{1}{q}}\left[ 
4(1-x)^{\frac{1-q}{q}}+(1+3x/2)^{\frac{1-q}{q}} \right]$ \\  
 & & & &   \\ \hline 
$N=6$  & $\left(\frac{1-x}{7}\right)\left(\frac{1-x}{6}\right)^{\frac{1-q}{q}}$  & $\left(\frac{1-x}{7}\right)\left(\frac{1+2x}{6}\right)^{\frac{1-q}{q}}$ & $\left(\frac{1+6x}{7}\right)\left(\frac{1+2x}{6}\right)^{\frac{1-q}{q}}$  &  $\left(\frac{1-x}{7}\right)\left(\frac{1}{6}\right)^{\frac{1}{q}}\left[ 
5(1-x)^{\frac{1-q}{q}}+(1+2x)^{\frac{1-q}{q}} \right]$ \\  
 & & & &   \\ \hline 
\end{tabular}}
\end{center}
\end{table}
Here too, there are only four distinct non-zero eigenvalues of the sandwiched matrix, two of which have $N-3$ and $2$-fold degeneracies respectively. As in the case of $\rho_N^{(W)}(x)$, here too the calculation of the eigenvalues for general $N$ is obtained by observing the trends of each column in Table ~2 for $N=3,\,4,\,5,\,6$. This leads to the four non-zero eigenvalues for all $N$ and they are given below;

\begin{eqnarray} 
\label{geneigghz}
\mu_1&=&\left(\frac{1-x}{N+1}\right)\left(\frac{1-x}{N}\right)^{\frac{1-q}{q}},\ \  \mbox{$(N-3)$-fold degenerate}; \nonumber \\
\mu_2&=&\left(\frac{1-x}{N+1}\right)\left(\frac{2+x(N-2)}{2N}\right)^{\frac{1-q}{q}}, \nonumber \\ 
\mu_3&=&\left(\frac{1+Nx}{N+1}\right)\left[\frac{2+x(N-2)}{2N}\right]^{\frac{1-q}{q}},  \\
\mu_4&=&\left(\frac{1-x}{N+1}\right) \left(\frac{1}{N}\right)^{\frac{1}{q}}\left[	\left(N-1\right)\left(1-x\right)^\frac{1-q}{q}+\left(1+\left(\frac{N}{2}-1\right)x\right)^{\frac{1-q}{q}}\right] \ \  \mbox{$2$-fold degenerate}; \nonumber 
\end{eqnarray}
The eigenvalues $\mu_i$ in Eq. (\ref{geneigghz}) allow us to find the value of $\tilde{D}^{T}_q(\rho^{(GHZ)}_N\vert\vert \rho_{B})=\frac{\sum_{i=1}^{N+1} \,\mu_i^q-1}{1-q}$ and the zero of the monotonically decreasing function $\tilde{D}^{T}_q(\rho^{(GHZ)}_N\vert\vert \rho_{B})$ is found to be at $x=\frac{2}{N^{2}+N+2}$ when $q\rightarrow \infty$. 
We thus have the $1: N-1$ separability range of the state $\rho_N^{(GHZ)}(x)$ using CSTRE criterion as    
\be
\label{sepghz}
0 \leq x \leq \frac{2}{N^{2}+N+2}
\ee 
for any $N\geq 3$.
We recall here that, in Ref. \cite{prabhu}, the separability range in the $1:N-1$ partition of the one parameter family of GHZ states was found using AR $q$-conditional entropy criterion and it matches exactly with Eq. (\ref{sepghz}). This is to be expected as, {\emph{the CSTRE criterion and AR criterion give the same results when the single qubit reduced density matrix turns out be a maximally mixed state thus commuting with its original density matrix~\cite{asanu}}}. Such a situation occurs in the case of one parameter family of noisy GHZ states~\cite{asanu} as the single qubit density matrix turns out to be  $I_A/2$, $I_A$ being the $2\times 2$ identity matrix.  Thus the results of CSTRE criterion match exactly with that of AR criterion in the case of one parameter family of noisy GHZ states. But {\emph{the difference between the CSTRE and AR criteria even in this case lies in the different modes of convergence of the parameter $x$ with the increase of $q$}}. In fact, $x$ converges slowly to the limit $2/(N^{2}+N+2)$ when CSTRE criterion is used whereas the convergence of $x$ is relatively fast for AR criterion. We have illustrated this feature for  $\rho_6^{(GHZ)}(x)$ in Fig. 3.

\begin{figure}[ht]
\begin{center}
\includegraphics* [width=2.4in,keepaspectratio]{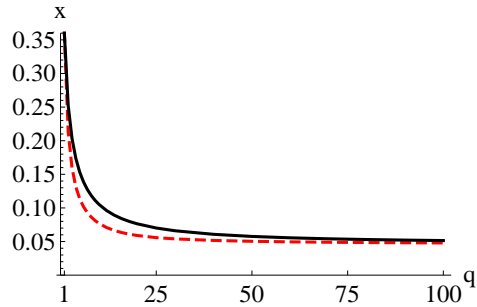} 
\caption{(Color Online) Implicit plots of $\tilde{D}^{T}_q(\rho^{(GHZ)}_6\vert\vert \rho_{B})=0$ as a function of $q$ (solid line) and Abe-Rajagopal $q$-conditional entropy $S_q^{T}(A\vert B)=0$ (dashed line) for $\rho_6^{(GHZ)}(x)$ in its $1:5$ partition. The relatively slow convergence of the parameter $x$ to $0.04545$ with the increase of $q$ in the case of CSTRE criterion is readily seen.  The quantities plotted are dimensionless.}  
\end{center}
\end{figure}

It is to be noticed that PPT criterion also gives the same $1: N-1$ separability range for $N=3,\,4,\,5,\,6$ for one-parameter family of GHZ states. Therefore, we can conjecture that Eq. (\ref{sepghz}) gives the PPT separability range in the $1:N-1$ partition of the one parameter family of noisy GHZ states $\rho_N^{(GHZ)}(x)$.  We have also observed that for large $N$ (macroscopic limit),  $x\approx \frac{2}{N^2}$ for $\rho_N^{(GHZ)}(x)$ and $x\approx \frac{\sqrt{2}-1}{N}$ for $\rho_N^{(W)}(x)$. Thus with the increase of $N$, the $1:N-1$ separability range decreases much faster for one parameter family of GHZ states than for one parameter family of W states.
 
It would be of interest to examine the state $\rho_N(x)$ (See Eq, (\ref{noisy})) when $\vert \Phi_N\rangle$ corresponds to the equal superposition of $W$ and its obverse counterpart ${\bar W}$: We have $\rho_N^{(W+{\bar W})}(x)$ to be  
\begin{eqnarray}
\label{wwbar}
\rho_N^{(W+{\bar W})}(x)&=&\left(\frac{1-x}{N+1}\right)P_N+ x\vert W{\bar W} \rangle_N \langle W{\bar W} \vert, \\ 
\vert W{\bar W}_N \rangle&=& \frac{1}{\sqrt{2}}\left(\vert W_N \rangle+\vert {\bar W}_N \rangle\right), \nonumber   
\end{eqnarray}
with $\vert {\bar W}_N \rangle=
\frac{1}{\sqrt{N}}\left(\vert 111\cdots 0\rangle+\vert 111\cdots 01\rangle+\cdots +\vert 
011\cdots 1\rangle \right)$ being the obverse counterpart of the  
W state $\vert  W_N \rangle=\frac{1}{\sqrt{N}}\left(\vert 000\cdots 1\rangle+\vert 
000\cdots 10\rangle+\cdots +\vert 100\cdots 0\rangle \right)$. In fact, the 
$3$-qubit GHZ state $\vert GHZ_3\rangle$ and $\vert W{\bar W}_3 \rangle$ are convertible into one another through Stochastic Local Operations and Classical Communications (SLOCC).  Both these states belong to the family of three distinct Majorana spinors whereas the W-state $\vert W_3\rangle$ belongs to the family of two distinct Majorana spinors\cite{maj}. In spite of belonging to the same SLOCC family, the entanglement features of $\vert GHZ_3\rangle$ and $\vert W{\bar W}_3 \rangle$ are shown to be quite different in Ref. \cite{maj}. Such a feature is also reflected in the symmetric noisy states containing these states. While the single qubit marginal of $\rho_3^{(GHZ)}(x)$ is maximally mixed thereby yielding the strictest $1:2$ separability range through AR-criterion itself, the corresponding $\rho_1\times I_{4}$ does not commute with $\rho_3^{(W+{\bar W})}(x)$ hence requiring CSTRE criterion for proper identification of its $1:2$ separability range. In fact, we have
\[
\rho_1=\frac{1}{6}\ba{cc} 3 & 2x \\ 2x & 3 \ea \Longrightarrow \rho_1\times I_{4}\ \ \mbox{does not commute with} \ \ \rho_3^{(W+{\bar W})}(x).
\]
On identifying the $1:2$ separability range of the symmetric state  $\rho_3^{(W+{\bar W})}(x)$ through AR-criterion, we find that it is given by  $(0,\,0.3333)$ which is evidently weaker compared to $(0,\,0.1896)$, the separability range obtained through CSTRE- as well as  PPT criteria. {\emph{But the $1:N-1$ separability range of the $N$-qubit state $\rho_N^{(W+{\bar W})}(x)$ where $N\geq 4$, carried out through a similar analysis as that for  $\rho_N^{(W)}(x)$, $\rho_N^{(GHZ)}(x)$ is found to be $0 \leq x \leq \frac{2}{N^{2}+N+2}$ for $N\geq 4$. It can be readily seen that this is identical to the $1:N-1$ separability range for the state $\rho_N^{(GHZ)}(x)$ (See Eq. (\ref{sepghz}))}. The AR-criterion is also found to give the same $1:N-1$ separability range $0 \leq x \leq \frac{2}{N^{2}+N+2}$ ($N\geq 4$) for $\rho_N^{(W+{\bar W})}(x)$. We have verified that the equivalence of the $1:N-1$ separability ranges for $\rho_N^{(W+{\bar W})}(x)$ through CSTRE- and AR-criteria when $N\geq 4$ is due to the {\emph{maximally mixed}} (hence commuting) nature of single qubit density matrix of $\rho_N^{(W+{\bar W})}(x)$ for $N\geq 4$.  Thus we can conclude that the $3$-qubit symmetric noisy state $\rho_3^{(W+{\bar W})}(x)$ stands out in showing different entanglement features than its higher qubit counterparts $\rho_N^{(W+{\bar W})}(x)$, $N\geq 4$. 

While the $1:N-1$ separability range of one parameter families of $N$-qubit states using CSTRE criterion matched exactly with that obtained through PPT criterion, similar conclusion cannot be drawn about separability ranges of other bipartitions such as $2:N-2$. For instance, the $2:2$ separability range of  
$\rho_4^{(W)}(x)$ using CSTRE criterion is found to be~\cite{asanu} $(0,\,0.2105)$ but PPT criterion yielded  $(0,\,0.0808)$ as the $2:2$ separability range of $\rho_4^{(W)}(x)$. Similarly, for  $\rho_4^{(GHZ)}(x)$ the $2:2$ separability ranges obtained through CSTRE and PPT criterion are found to be $(0,\,0.2105)$, $(0,\,0.0625)$ respectively. But an investigation into  bipartitions such as $3:N-3$ and so on may yield results which are either identical, weaker or stricter than that through PPT criterion. It is also to be noted here that PPT and CSTRE separability critera are of entirely different origins and there is no reason to expect that the separability ranges obtained through them match with each other in all bipartitions of an $N$-qubit state. It is indeed surprising that the $1:N-1$ separability ranges of one parameter families of states that we have investigated here, obtained using PPT and CSTRE crtierion, matched exactly with each other whereas the separability ranges in other bipartitions may not do so. It is worth recalling here that CSTRE criterion is non-spectral in nature and is able to distinguish between isospectral states~\cite{asanu}. In view of the fact that no spectral criteria can detect entangled states with positive partial transpose~\cite{spbe}, the so-called bound entangled states, it is worth examining whether CSTRE criterion can detect bound entangled states by providing a separability range stricter than the PPT criterion. Having seen that CSTRE criterion provides separability ranges either identical or weaker in comparison with PPT criterion,  whether it can provide a separability range stricter than that through PPT criterion in {\emph {atleast }}some bipartitions of the $N$-qubit state still remains an open question.    

It would be of interest to examine completely random multi-qubit states and analyze the separability ranges in different bipartitions, obtained through CSTRE and PPT criteria. While a numerical investigation on the set of all bipartite mixed states has revealed~\cite{jb} that PPT criterion is superior to AR criterion, a comprehensive numerical survey on the PPT as well as CSTRE separability ranges in different bipartitions of random states is warranted in order to identify the hierarchy between CSTRE and PPT criteria. Such a numerical survey, on the same lines as in  Ref. ~\cite{jb}, will also help in strengthening the results of the present article.                 
\subsection{Further illustrations and future directions}
Having used the conditional version of sandwiched relative Tsallis entropy to find the separability range in symmetric one parameter family of noisy $W$-, GHZ and $W\bar{W}$ states, we have also examined the utility of conditional version of sandwiched R\'{e}nyi relative entropy in finding whether a bipartite state is entangled or not. 
We have found that both Tsallis and R\'{e}nyi entropies play the same role in the detection of bipartite entanglement in a quantum state.   

The conditional version of sandwiched R\'{e}nyi relative entropy is given by 
\be
\label{reny}
\tilde{D}^{R}_q(\rho_{AB}||\rho_B)=\frac{\log\left[\tilde{Q}_q(\rho_{AB}||\rho_B)\right]}{1-q}
\ee
where $\tilde{Q}_q(\rho_{AB}||\rho_B)$ is as shown in Eq. (\ref{qab}). 
We have evaluated the range of the parameter $x$ where  $\tilde{D}^{R}_q(\rho_{AB}||\rho_B)$ is greater than zero and observe that the same result as obtained through CSTRE is obtained for both $\rho_N^{(W)}(x)$, $\rho_N^{(GHZ)}(x)$. 
This implies that R\'{e}nyi entropy which is additive plays the same role as the non-additive Tsallis entropy in the identification of entanglement in the symmetric one-parameter families of N-qubit states. One can expect that this feature remains true for all bipartite states and one can either choose $\tilde{D}^{R}_q(\rho_{AB}||\rho_B)$ (Eq. (\ref{reny})) or $\tilde{D}^{T}_q(\rho_{AB}||\rho_B)$ (Eqs. (\ref{cstre1}), (\ref{qab})) for detecting bipartite entanglement.  

At this juncture, we notice that in Ref. \cite{renyimax}, a conditional version of R\'{e}nyi relative entropy is defined by maximizing over the marginal state $\rho_B$. While the results obtainable through such a maximization over $\rho_B$ are of interest, in view of the fact that it is operationally difficult to identify $\rho_B$ that maximizes the conditional entropy (either R\'{e}nyi or Tsallis), our analysis here is restricted to the case of the actual marginal $\rho_B$ of the bipartite state $\rho_{AB}$. It would be interesting to examine the consequences of maximization over marginals in the detection of entanglement using conditional generalized entropies (R\'{e}nyi or Tsallis) and at present it remains an open problem.  

We wish to mention here that the {\emph {applicability of CSTRE is not restricted to the symmetric one-parameter family of noisy $W$-, GHZ, $W\bar{W}$ states}}. In fact, the CSTRE criterion is applicable for identifying any bipartite entangled state and one can use it for obtaining the separability ranges in chosen bipartitions of several one-parameter, two-parameter families of states including X states, cluster/graph states.  An example of the use of CSTRE in identifying entanglement in an isospectral family of 2-qubit X states is illustrated in Ref.~\cite{asanu}. Also the applicability of CSTRE is not restricted to composite quantum states with two level systems (qubits) alone and it encompasses mixed composite states with qudits also. For instance, let us consider the one parameter family of $3\times 3$ isotropic state given by~\cite{dep} 
\[
\rho_{ab}(x)=\left(\frac{1-x}{8}\right)I_{9}+\left(\frac{9x-1}{8}\right) \vert \Phi \rangle \langle \Phi \vert, \ \ \vert \Phi\rangle=\frac{1}{\sqrt{3}}\left(\vert 00 \rangle+\vert 11 \rangle+\vert 22 \rangle\right),  
\]
with $0\leq x \leq 1$, $I_9$ is $3^2\times 3^2$ identity matrix, $\vert 0\rangle=(1,\,0,\,0)$, $\vert 1\rangle=(0,\,1,\,0)$, $\vert 2\rangle=(0,\,0,\,1)$ are the basis states in the qutrit space. The single qutrit reduced subsystems $\rho_a$, $\rho_b$ turn out to be $I_3/3$ thereby commuting with $\rho_{ab}$. The CSTRE criterion thus reduces to AR-criterion and on evaluating $\tilde{D}^{T}_q(\rho_{ab}(x)||\rho_a)$ ($\equiv \tilde{D}^{T}_q(\rho_{ab}(x)||\rho_b)$) we obtain the separability range of the state to be $(0,\, 1/3)$ which matches exactly with that obtained through PPT criterion. 

The CSTRE criterion is also useful in identifying entanglement in $d_1\times d_2$ dimensional states as can be seen through the example of a qubit-qutrit ($2\times 3$) X state~\cite{dep,esd10}. The state $\rho_{ab}$ is given by~\cite{dep,esd10} 
\[
\rho^X_{ab}(x)=\frac{1}{8} \ba{cccccc} 2 & 0 & 0 & 0 & 0 & 8x \\ 0 & 1 & 0 & 0 & 0 & 0 \\ 0 & 0 & 1 & 0 & 0 & 0 \\ 
0 & 0 & 0 & 1 & 0 & 0 \\ 0 & 0 & 0 & 0 & 1 & 0  \\ 8x & 0 & 0 & 0 & 0 & 2 \ea, \ \ \ 0\leq x \leq 1/4
\]  
and its subsystem $\rho_b=\mbox{Tr}_a\,\rho_{ab}$ corresponding to the qutrit is found to be $\frac{1}{8}\mbox{diag}\,(3,\,2,\,3)$. On evaluating $\tilde{D}^{T}_q(\rho^X_{ab}(x)||\rho_b)$, the separability range for the state  $\rho^X_{ab}(x)$ is obtained to be $(0,\,1/8)$ quite in agreement with the range obtained through PPT criterion.  Observe that the other marginal, the single qubit density matrix $\rho_a$ turns out to be maximally mixed 
with $\rho_a=I_2/2$ thus implying the equivalence of CSTRE with AR-criterion when worked with this marginal. But $\tilde{D}^{T}_q(\rho^X_{ab}(x)||\rho_a)\geq 0$ for all values of $x\in (0,\,1/4)$ thus failing to capture the entanglement in the state.  This example illustrates the need for suitable choice of marginals in making effective use of CSTRE criterion.     

One can also use the CSTRE criterion to examine separability in non-symmetric analogues of the family of states $\rho_N(x)$ (See Eq. (\ref{noisy})). The states 
$\rho_N(x)$ considered in Eq. (\ref{noisy}) are {\emph{noisy states belonging to the symmetric subspace}} because the white noise represented by $P_N=\sum_{M=-\frac{N}{2}}^{\frac{N}{2}}\,\left\vert \frac{N}{2},\,M \right\rangle \left\langle \frac{N}{2},\,M\right\vert$ is the identity operator in the $N+1$ dimensional symmetric subspace of collective angular momentum $j=N/2$.  Their non-symmetric counterparts are given by 
\be
\label{ft1}
\rho^{ns}_N(x)=\left(\frac{1-x}{2^N}\right)I_{2^N}+ x\vert \Psi_N \rangle \langle \Psi_N \vert, \ \ I_{2^N}\equiv\,2^N \times 2^N \ \mbox{identity matrix}
\ee
with $I_{2^N}$ representing the white noise in $2^N$ dimensional space and $\vert \Psi_N \rangle$ are either symmetric or non-symmetric $N$-qubit states. The $N$-qudit noisy states 
\be
\label{ft2}
\rho^{(d)}_N(x)=\left(\frac{1-x}{d^N}\right)I_{d^N}+ x\vert \Psi^{(d)}_N \rangle \langle \Psi^{(d)}_N \vert; \ \ I_{d^N}\equiv \,d^N \times d^N \mbox{identity matrix}
\ee 
can also be examined using CSTRE criterion. The applicability of AR-criterion to $\rho^{(d)}_N(x)$ when $\vert \Psi^{(d)}_N\rangle$ corresponds to the $d$-dimensional analogue of GHZ state has been carried in Ref. \cite{sabe} and it was shown that the AR-criterion yields the strictest possible $1:N-1$ separability range that matches with the range obtained through PPT criterion~\cite{sabe}. It would be of interest to analyze the bipartite separability range in $\rho^{(d)}_N(x)$ when 
$\vert \Psi^{(d)}_N\rangle$ is the $d$-dimensional analogue of $W$ state and examine whether CSTRE criterion fares better than AR-criterion. In fact, one parameter family of noisy states (symmetric or otherwise) involving W states yield better $1:N-1$ separability range through CSTRE as their single-qubit reduced density matrices are not maximally mixed, thus making them suitable for an analysis through CSTRE criterion, a non-commuting generalization of the AR-criterion. For instance, our preliminary investigations on the $3$-qubit noisy state $\rho^{ns}_3(x)$ (See Eq. (\ref{ft1})) with $\vert \Psi_3\rangle$ being the $3$-qubit W state yields a $1:2$ separability range $(0,\,0.2096)$ using CSTRE criterion which is stronger in comparison with $(0,\,0.2727)$  obtained through AR-criterion. 
The generalization of this result to $N$-qubit/qudit noisy states $\rho^{ns}_N(x)$/$\rho^{(d)}_N(x)$ both involving W states/generalized W states will be presented in a forthcoming work.

\section{Conclusion} 
\label{sec4}
In this work, we have shown that negative values of the conditional version of sandwiched Tsallis relative entropy necessarily imply quantum entanglement in a bipartite state. Using this result and considering the limit $q\rightarrow \infty$ in Tsallis entropy, we have obtained the separability range of the symmetric one-parameter family of noisy $N$-qubit $W$, GHZ, $W\bar{W}$ states in their $1:N-1$ partition. For the one-parameter family of noisy W-states we have shown that the CSTRE criterion provides a stricter $1:N-1$ separability range when compared to that obtained through AR $q$-conditional entropy approach. The non-commutativity of the single qubit marginal density matrix with the original density matrix of the noisy $N$-qubit W states is seen to be the reason behind the supremacy of CSTRE criterion over AR criterion. The $1:N-1$ separability range, obtained using CSTRE criterion, for the one-parameter family of noisy GHZ states matches with that through AR criterion. This is due to the maximally mixed, thereby commuting nature of the single qubit density matrix  with the original density matrix in the symmetric one parameter family of noisy $N$-qubit GHZ states. 
We have thus illustrated CSTRE criterion as a non-commuting generalization of the AR criterion and its equivalence with the results of AR criterion in the commuting cases, wherein the marginals are maximally mixed.  In view of the fact that the $1:N-1$ separability ranges through CSTRE and PPT criterion match with each other, our work has provided the PPT separability range also for the one-parameter families of states considered here, in their $1:N-1$ partition. We have also indicated that CSTRE separability ranges in bipartitions other than $1:N-1$ may not match with that through PPT criterion.  Our analysis, using AR- and CSTRE criteria, of the one parameter family of noisy state involving the state $\vert W{\bar W}_N\rangle$, an equal superposition of $W$, obverse $W$ states, has revealed an interesting feature that the $3$-qubit state of this family shows a different entanglement feature than its higher qubit counterparts. 

We have given further illustrations on the applicability of CSTRE criterion to $d\times d$ as well as $d_1 \times d_2$ states by considering a two qutrit isotropic state and a qubit-qutrit X state. We have quoted the results of our preliminary investigation on the separability ranges in the non-symmetric one-parameter family of noisy $N$-qubit W states and indicated the need for further exploration through the CSTRE criterion.   

It would be of interest to explore the separability ranges of one parameter families of mixed $N$-qubit states in their different bipartitions using CSTRE criterion and explore how non-commutativity aspect plays a role in finding stricter separability ranges compared to the existing separability criteria. In view of the non-spectral nature of CSTRE criterion~\cite{asanu}, whether it can fare better than the PPT criterion and can detect bound entangled states remains an open problem as of now.        

The usefulness of CSTRE criterion in finding the separability ranges in one-parameter, two-parameter families of symmetric/non-symmetric $N$-qubit/qudit quantum states will be the content of our forthcoming work. A numerical investigation of separability ranges using CSTRE and PPT criteria in different bipartitions of a random mixed state is another work that we wish to take up in the future. 

\section*{Acknowledgment}
Anantha S Nayak acknowledges the support of Department of Science and
Technology (DST), Govt. of India through the award
of INSPIRE fellowship.  

\end{document}